\documentclass[aps,twocolumn,superscriptaddress,showpacs,floats]{revtex4}
\usepackage{amsmath}
\usepackage{amsfonts}
\usepackage{amssymb}
\usepackage{euscript}
\usepackage{bm}

\begin{document}

\title{Vortex--Phonon Interaction}

\author{Evgeny Kozik}
\affiliation{Department of Physics, University of Massachusetts,
Amherst, Massachusetts 01003, USA}
\author{Boris Svistunov}
\affiliation{Department of Physics, University of Massachusetts,
Amherst, Massachusetts 01003, USA} \affiliation{Russian Research
Center ``Kurchatov Institute'', 123182 Moscow, Russia}

\begin{abstract}
Kelvin waves (kelvons)---helical waves on quantized vortex
lines---are the normal modes of vortices in a superfluid. At zero
temperature, the only dissipative channel of vortex dynamics is
phonon emission. Starting with the hydrodynamic action, we derive
the Hamiltonian of vortex-phonon interaction, thereby reducing the
problem of the interaction of Kelvin waves with sound to inelastic
elementary excitation scattering. On the basis of this formalism,
we calculate the rate of sound radiation by superfluid turbulence
at zero temperature and estimate the value of short-wavelength
cutoff of the turbulence spectrum.
\end{abstract}


%
%
%
%

\maketitle

Since the early study of phonon scattering by vortex lines
\cite{Fetter_scattering}, much interest has been attracted by the
problems of the interaction of Kelvin waves with
density-distortion modes \cite{Epstein_Baym, Vinen2000, Vinen2001,
Bretin, Mizushima, Stoof}. In helium, Kelvin waves are a
fundamental ingredient of the evolution of superfluid turbulence
\cite{Donnelly, Cambridge, Davis, Sv_95, Vinen2000, Vinen2001,
SvK_2004}. A Kolmogorov-type cascade of Kelvin waves has been
argued  \cite{Sv_95, Vinen2000, SvK_2004} to be responsible for
the decay of superfluid turbulence at zero temperature
\cite{Davis}. As proposed by Vinen \cite{Vinen2000, Vinen2001},
the sound radiation by short-wavelength kelvons leads to a
dissipative cutoff of the Kolmogorov energy flux. In neutron
stars, the excitation of Kelvin waves due to the interaction with
the nuclei in the solid crust is suggested to be the main
mechanism of pulsar glitches \cite{Epstein_Baym}. Not long ago,
quantized vortices were created in atomic Bose-Einstein condensate
(BEC) \cite{Cornell}. Since then, nonlinear kelvon dynamics have
become an attractive topic in the field of ultra-cold gases
\cite{Bretin, Mizushima, Stoof, Smith}. Recent experiment in
gaseous BEC by Bretin \textit{et. al.} \cite{Bretin} and the
subsequent numerical simulation by Mizushima \textit{et. al.}
\cite{Mizushima} describe the excitation of Kelvin waves by the
quadrupole modes. In their theoretical study, Martikainen and
Stoof \cite{Stoof} obtained the interaction Hamiltonian of kelvons
with the quadrupole modes in a model of a vortex line in a stack
of two-dimensional BEC.

We develop a systematic approach to the problem of interaction of
phonons with vortices in the hydrodynamic regime, i.e. when any
physical length scale is much larger than the vortex core size
$a_0$, which allows one to describe vortices as geometrical lines
\cite{Donnelly}. We derive the interaction Hamiltonian basing the
analysis on the small parameter $\beta = a_0 \tilde{k} \ll 1$,
where $\tilde{k}$ is the largest wave number among kelvons and
phonons. To employ the transparent description in terms of the
normal modes, we confine ourselves to the case of weak
nonlinearity. For kelvons this implies that the amplitudes $b_k$
of the Kelvin waves of the typical wavelength $\lambda \sim
k^{-1}$ are much smaller than $\lambda$, which is expressed by the
small parameter $\alpha_k = b_k k \ll 1$. For phonons this
requires that $\eta \ll n$, where $\eta$ is the number density
fluctuation in a sound wave and $n$ is the average number density.
The obtained result allows us to rigorously describe the radiation
of sound by kelvons, which we apply to the problem of superfluid
turbulence decay at zero temperature. We should mention that, in
superfluid turbulence, kelvons are actually superimposed on vortex
kinks with curvature radii much larger than kelvon wavelengths
\cite{Sv_95}. We shall neglect such large-scale curvature and
address this issue in the end of the paper.

The condition $a_0 k \sim \beta \ll 1$ implies that the typical
vortex line velocities are much smaller than the speed of sound.
Along with $\eta \ll n$ it leads to the fact that the
vortex-phonon coupling contributes only small corrections to the
dynamics of the non-interacting vortex and phonon subsystems.
Therefore, a perturbative approach is applicable, provided the
interaction energy is written in terms of the canonical variables.
Normally, the form of the canonical variables comes from the
solution of the interaction-free dynamics. However, when studying
vortices separate from phonons, one naturally neglects the
compressibility of the fluid \cite{Donnelly}, since finite
compressibility leads only to higher-order ``relativistic''
corrections. As a result, the dynamics of vortices are described
by the Hamiltonian, written in terms of the geometrical
configuration of the vortex lines \cite{Sv_95}. When the vortices
are absent, the phonon modes come from the bilinear Hamiltonian
for the density fluctuation $\eta(\mathbf{r}, t)$ and the phase
field $\varphi(\mathbf{r}, t)$, which determines the velocity in
the density wave (see, e.g., \cite{LLStatMech2}). If finite
compressibility of a superfluid is taken into account in order to
join the subsystems, the positions of vortices and the fields
$\eta(\mathbf{r}, t), \varphi(\mathbf{r}, t)$ are no longer the
sets of canonical variables because of the variable-mixing term in
the Lagrangian. Introducing the interaction, one has to
simultaneously reconsider the canonical variables.

The small parameters allow us to obtain an asymptotic expansion of
the canonical variables by means of a systematic iterative
procedure. Physically, the procedure restores the retardation in
the adjustment of the superfluid velocity field to the evolving
vortex configuration. It qualitatively changes the structure of
the Hamiltonian with respect to the terms responsible for the
radiation of sound and the relativistic corrections to the vortex
dynamics.

Long-wave superfluid dynamics at zero temperature are described by
the Popov's hydrodynamic action \cite{Popov}:
\begin{gather}
S=\int \! \! \! dt \, d^3 r \left[-(n+\eta)\dot{\Phi}
 - \frac{(n + \eta)}{2m_0}
|\nabla \Phi |^2 - \frac{1}{2\varkappa} \eta^2 \right].
\label{Action}
\end{gather}
Here the spatial integral is taken over the macroscopic fluid
volume, $\Phi(\mathbf{r}, t)$ is the phase field, which determines
the velocity according to $\mathbf{v}=(1/m_0)\nabla \Phi$ ($\hbar
= 1$), $m_0$ is the mass of a particle, $\varkappa$ is the
compressibility, the dot denotes the derivative with respect to
time, and the vortex core radius is given by $a_0 \sim
\sqrt{\varkappa/n m_0}$.

The phase $\Phi$ is non-single-valued and contains topological
defects, the vortex lines. The velocity circulation around each
vortex line is quantized:
\begin{equation}
\oint \nabla \Phi(\mathbf{r}) \cdot d \mathbf{r} = 2 \pi.
\label{circulation}
\end{equation}
The defects can be separated from the regular contribution:
\begin{equation}
\Phi=\Phi_0 + \varphi, \label{Phi_0_plus_phi}
\end{equation}
where $\Phi_0$ is non-single-valued and satisfies
(\ref{circulation}), while $\varphi$ is regular and $\nabla
\varphi$ is circulation-free. The standard decomposition into
vortices and phonons is done by introducing an additional
constraint that to the zeroth approximation eliminates the
coupling between $\Phi_0$ and $\varphi$ in the Hamiltonian, namely
\begin{equation}
\Delta \Phi_0(\mathbf{r})=0. \label{Laplacian}
\end{equation}
(Physically, this parametrization is suggested by the velocity
potential of an incompressible fluid.) With
Eqs.~(\ref{Phi_0_plus_phi}),(\ref{Laplacian}) the Lagrangian
becomes
\begin{equation}
L= \int \! \! \! d^3 r \left[- n \dot{\Phi}_0 - \eta\dot{\varphi}
- \eta \dot{\Phi}_0 \right] - H, \label{Lagrangian}
\end{equation}
where $H=H_{\mathrm{vor}} + H_{\mathrm{ph}} + H'_{\mathrm{int}}$,
\begin{gather}
H_{\mathrm{vor}} = \frac{n}{2m_0} \int \! \! \! d^3 r \;
\bigl| \nabla \Phi_0 \bigr|^2, \label{H_vor} \\
H_{\mathrm{ph}} = \int \! \! \! d^3 r \left[ \frac{n}{2m_0} \bigl|
\nabla \varphi \bigr|^2 + \frac{1}{2 \varkappa}\eta^2
\right], \label{H_ph} \\
H'_{\mathrm{int}} = \frac{1}{2m_0} \int \! \! \! d^3 r \Bigl[ \;
\eta \bigl|\nabla \Phi_0\bigr|^2 + 2 \eta \nabla \varphi \cdot
\nabla \Phi_0 \Bigr]. \label{H_int}
\end{gather}
The coupling between the vortex variable, $\Phi_0$, and the
density waves, $\{\eta, \varphi\}$, is determined by
$H'_{\mathrm{int}}$ and the time derivative term $\int \! d^3 r \,
\eta \dot{\Phi}_0$, both being first-order corrections to the
non-interacting parts. Following standard perturbative procedure,
we first neglect this coupling to find the non-interacting normal
modes. The vortex part of the Lagrangian is then given by
$L_{\mathrm{vor}}= - n \int \! d^3 r \; \dot{\Phi}_0 -
H_{\mathrm{vor}}$. For the sake of simplicity, from now on we
consider a solitary vortex line; the generalization is
straightforward. Let the two-dimensional vector
$\boldsymbol{\rho}_0(z_0) = \bigl(x_0(z_0) , y_0(z_0), 0\bigr)$
describe the position of the vortex line in the plane $z=z_0$ of a
Cartesian coordinate system, where the $z$-direction is chosen
along the vortex line. The field $\Phi_0$ is a functional of
$\boldsymbol{\rho}_0(z)$, hence
\begin{equation}
\int \! \! \! d^3 r \; \dot{\Phi}_0 = \int \! \! \! d z \;
\dot{\boldsymbol{\rho}}_0 \cdot \frac{\delta}{\delta
\boldsymbol{\rho}_0} \int \! \! \! d^3 r \; \Phi_0. \label{int1}
\end{equation}
To obtain $\int \! d^3 r \; \dot{\Phi}_0$, it is sufficient to
calculate the integral $\int \!  d^3 r \; \delta \Phi_0$, where
$\delta \Phi_0$ is the variation of the phase field due to the
distortion of the vortex line by $\delta \boldsymbol{\rho}_0(z)$.
Using the identity $\Delta(\boldsymbol{\rho}^2)=4$, with
$\boldsymbol{\rho}=(x, y, 0)$, and Eq.~(\ref{Laplacian}), obtain
\begin{equation}
\int \! \! \! d^3 r \; \delta \Phi_0 = \frac{1}{4} \int \! \! \!
d^3r \; \nabla \cdot \left[ \; \delta \Phi_0 \nabla
(\boldsymbol{\rho}^2) \; \right]. \label{int2}
\end{equation}
The variation $\delta \Phi_0 (\mathbf{r})$ can be viewed as being
produced by two vortex lines with opposite circulation quanta and
separated by $\delta \boldsymbol{\rho}_0(z)$. In view of
Eq.~(\ref{circulation}) the field $\delta \Phi_0(\mathbf{r})$
experiences a jump of $2\pi$ across the surface $\mathcal{S}$ that
extends between these vortex lines along the vector field $\delta
\boldsymbol{\rho}_0(z)$, therefore the integration volume must
have a cut along $\mathcal{S}$. Applying the Gauss theorem, we
rewrite Eq.~(\ref{int2}) as the surface integral $(1/2)
\oint_{\mathcal{S}} \delta \Phi_0 \; (\boldsymbol{\rho} \cdot d
\mathbf{S} )$ yielding
\begin{equation}
\int \!\!\! d^3r \; \dot{\Phi}_0 = \pi \int \!\!\! d z \;
 [ \mathbf{\hat{z}} \times \boldsymbol{\rho}_0(z) ] \cdot
\dot{\boldsymbol{\rho}}_0(z) . \label{int4}
\end{equation}
Introducing the complex variable $w(z)= \sqrt{n m_0 \kappa/2} \;
\bigl[ x(z)+iy(z) \bigr]$, where $\kappa = 2\pi/m_0$ is the
velocity circulation quantum, obtain
\begin{equation}
L_{\mathrm{vor}}= \frac{1}{2} \int \!\!\! dz \; \bigl[ i w^*
\dot{w} - i \dot{w}^* w \bigr] - H_{\mathrm{vor}}[w, w^*],
\label{L_vor}
\end{equation}
which implies that $w(z)$ and $w^*(z)$ are the canonical variables
with respect to $L_{\mathrm{vor}}$. The energy (\ref{H_vor}),
rewritten in terms of $w(z)$ and $w^*(z)$ gives the vortex
Hamiltonian \cite{Sv_95}. In this paper we keep only the bilinear
term of the expanded with respect to $\alpha_k \ll 1$ Hamiltonian,
the remaining terms determine the interaction of kelvons
\cite{SvK_2004}. The result is \cite{Fetter_quant}
\begin{equation}
 H_{\mathrm{vor}} \approx \sum_k \varepsilon_k a^{\dagger}_k a_k \;,
\;\; \varepsilon_k = (\kappa/4\pi) \ln (1/k a_0) k^2,
\label{kelvons}
\end{equation}
where $a_k$ and $a^{\dagger}_k$ are the kelvon creation and
annihilation operators, obtained, assuming the periodicity along
$z$, by the Fourier transforms $a_k=L^{-1/2} \int w(z) \exp (-i k
z) \, dz$, where $L$ is the system size in the $z$-direction, and
$w(z)$ is understood as a quantum field.

The sound waves are described by the Lagrangian $L_{\mathrm{ph}}=
\int \! d^3 r \left(- \eta\dot{\varphi}\right) -
H_{\mathrm{ph}}[\eta, \varphi] $, with $H_{\mathrm{ph}}$ given by
(\ref{H_ph}). We assume that the system is contained in a cylinder
of radius $R$ with the symmetry axis along the $z$-direction and
that the system is periodic along $z$ with the period $L$. In the
cylindrical geometry, the phonon fields $\eta(r, \theta, z)$,
$\varphi(r, \theta, z)$ are parametrized by phonon creation and
annihilation operators $c_s$, $c^{\dagger}_s$ as
\begin{gather}
\eta = \sum_s \sqrt{\,\omega_s \, \varkappa\,/\,2} \; \left[ \;
\chi_s \, c_s + \chi^*_s\, c^{\dagger}_s \; \right], \notag
\\ \varphi = - i \sum_s \sqrt{1\, /\, 2\,\omega_s \, \varkappa} \; \left[ \;
\chi_s \, c_s - \chi^*_s\, c^{\dagger}_s \; \right],
\label{phonon_fields} \\
\chi_s=\chi_s (r, \theta, z) = \mathcal{R}_{m q_r} \!(r) \; Y_m
\!(\theta) \; \mathcal{Z}_{q_z}\!(z), \notag
\end{gather}
where $\mathcal{R}_{m q_r}(r)= (\pi q_r/R)^{1/2} J_m(q_r r)$, $Y_m
\! (\theta) = (2\pi)^{-1/2}  \exp(i m \theta)$,
$\mathcal{Z}_{q_z}\!(z) = L^{-1/2}  \exp ( i q_z  z )$, $s$ stands
for $\{ q_r, m, q_z \}$, and $J_m(x)$ are the Bessel functions of
the first kind. The phonon Hamiltonian then reads
\begin{equation}
H_{\mathrm{ph}}=\sum_s \omega_s c^{\dagger}_s c_s, \;\;\;
\omega_s= c \, q,\label{phonons}
\end{equation}
with $q= \sqrt{q_r^2 + q_z^2}$ and $c=\sqrt{n/\varkappa m_0}$.

Now we address the coupling between phonons and vortices. In terms
of the obtained variables, the Lagrangian (\ref{Lagrangian}) takes
on the form
\begin{equation}
L = \sum_k i \, \dot{a}_k a^{\dagger}_k + \sum_s i \, \dot{c}_s
c^{\dagger}_s - T  - H , \label{Lag_old_variables}
\end{equation}
where $T = \int \! d^3r \; \eta\dot{\Phi}_0 \equiv
T\{a_k,\dot{a}_k, a^{\dagger}_k,\dot{a}^{\dagger}_k, \;
c_s,\dot{c}_s, c^{\dagger}_s, \dot{c}^{\dagger}_s\}$, and
$H=H_{\mathrm{vor}}+H_{\mathrm{ph}}+H'_{\mathrm{int}}$. The
coupling term $T$ plays a special role in the Lagrangian
(\ref{Lag_old_variables}). This term is linear in time derivatives
of the variables and thus can not contribute to the energy in
accordance with the Lagrangian formalism. Moreover, because of the
time derivatives in $T$, the equations of motion in terms of
$\{a_k, a^{\dagger}_k\}, \{c_s, c^{\dagger}_s\}$ take on a
non-Hamiltonian form. This implies that the chosen variables
become non-canonical in the presence of the interaction, and
therefore the total energy, $H$, \textit{in these variables} can
not be identified with the Hamiltonian.

There must exist such a variable transformation $\{a_k,
a^{\dagger}_k\}, \{c_s, c^{\dagger}_s\} \rightarrow \{\tilde{a}_k,
\tilde{a}^{\dagger}_k\}, \{\tilde{c}_s, \tilde{c}^{\dagger}_s\}$
that restores the canonical form of the Lagrangian, $L = \sum_k i
\, \dot{\tilde{a}}_k \tilde{a}^{\dagger}_k + \sum_s i \,
\dot{\tilde{c}}_s \tilde{c}^{\dagger}_s  - H \{\tilde{a}_k,
\tilde{a}^{\dagger}_k,\tilde{c}_s, \tilde{c}^{\dagger}_s\}$. The
canonical variables are obtained by the following iterative
procedure. The term $T$ is expanded with respect to $\alpha_k \ll
1$, $\beta \ll 1$ and $\eta \ll n$ yielding
$T=T^{(1)}+T^{(2)}+\cdots$. Then the variables are adjusted by
$a_k \rightarrow a_k + a^{(1)}_k$, $c_s \rightarrow c_s +
c^{(1)}_s$, where $a^{(1)}_k ( \{a_k, a^{\dagger}_k, c_s,
c^{\dagger}_s\})$ and $c^{(1)}_s (\{a_k, a^{\dagger}_k, c_s,
c^{\dagger}_s\})$ are chosen to eliminate the term $T^{(1)}$ in
(\ref{Lag_old_variables}). As a result, $T \rightarrow 0 +
T'^{(2)} + \cdots$, where the prime means that the structure of
the remaining terms has changed. At the next step, $T'^{(2)}$ is
eliminated by $a_k \rightarrow a_k + a^{(2)}_k$, $c_s \rightarrow
c_s + c^{(2)}_s$ and so on. By construction, the canonical
variables are given by $\tilde{a}_k = a_k + a^{(1)}_k + a^{(2)}_k
+ \cdots$, $\tilde{c}_s = c_s + c^{(1)}_s + c^{(2)}_s + \cdots$,
and likewise for the conjugates. In practice, only the first few
terms are enough, as the rest ones give just higher-order
corrections.

The explicit expression for $T$ is obtained following the steps of
the derivation of Eq.~(\ref{int4}). The only difference here is
that the role of the auxiliary function $\boldsymbol{\rho}^2$ is
played by $Q$, defined by $\Delta Q (\mathbf{r}) \equiv
\eta(\mathbf{r})$:
\begin{equation}
T = 2 \pi  \int \!\!\! d z \;  [ \mathbf{\hat{z}} \times \nabla
Q(\boldsymbol{\rho}_0(z), z) ] \cdot \dot{\boldsymbol{\rho}}_0(z)
. \label{T1}
\end{equation}

In accordance with $q_r \rho_0 \sim \beta k \rho_0 \ll 1$, we
expand the radial functions, retaining only the greatest term for
each particular angular momentum $m$. Switching to the cylindrical
coordinates, $\boldsymbol{\rho}_0 = (\rho_0 \, \cos \gamma, \;
\rho_0 \, \sin \gamma, \; 0 )$, and noticing that
\begin{multline}
\left[ \dot{\gamma} \, \rho_0 \, \frac{\partial}{\partial r} -
\frac{\dot{\rho}_0}{\rho_0}\, \frac{\partial }{\partial \theta}
\right]\mathcal{R}_{m q_r}\!(r)Y_m \! (\theta) \Bigl|_{r=\rho_0,
\theta=\gamma} \\
\propto \left\{ \begin{array}{cc}   d( w^{m})/dt ,& m\geq0
\\ d( w^{*|m|})/dt, & m<0
\end{array} \right. , \label{w.z}
\end{multline}
obtain
\begin{multline}
T \approx \sum_{s, k_1 \ldots k_m} \Bigl[ \;
 - iA_{s,k_1\ldots k_m}\, c_s \; \frac{d}{dt}(a_{k_1}\ldots
a_{k_m}) \\ - iB_{s,k_1\ldots k_m}\, c_s \;
\frac{d}{dt}(a^{\dagger}_{k_1}\ldots a^{\dagger}_{k_m})\; \Bigr] +
\mathrm{H.c.}, \label{T_final}
\end{multline}
where the sum is over all $s$ with $m\neq0$ and
\begin{gather}
A_{s,k_1\ldots k_m}= -\Theta(m) \, A_s \, \delta_{k_1+\ldots +k_m, -q_z}, \notag \\
B_{s,k_1\ldots k_m}= (-1)^{|m|} \Theta(-m -1) \, A_s \,
\delta_{k_1+\ldots +k_m, q_z}, \notag \\ A_s=\frac{\sqrt{a_0
q}}{2^{|m|/2+1}\; |m|!} \, \frac{n^{\frac{1-|m|}{2}} (m_0
\kappa)^{\frac{2-|m|}{2}} \, q_r^{|m| + \frac{1}{2}} \,
q^{-2}}{L^{(|m|-1)/2} \; R^{1/2}}, \label{vertex}
\end{gather}
where $\Theta(m) = \left\{
\begin{array}{cc}   1,& m\geq0
\\ 0, & m<0
\end{array} \right.$. Thus,
\begin{equation}
a_k=\tilde{a}_k \label{transf.kelvon}
\end{equation}
(we omitted the terms that do not contain phonon operators and
thus result only in relativistic corrections to the kelvon
spectrum and kelvon-kelvon interactions),
\begin{multline}
c_s=\tilde{c}_s + (1-\delta_{m,0}) \sum_{k_1 \ldots k_m}
\Bigl[A_{s,k_1\ldots k_m}\tilde{a}^{\dagger}_{k_1}\ldots
\tilde{a}^{\dagger}_{k_m}  \\
+B_{s,k_1\ldots k_m} \tilde{a}_{k_1}\ldots \tilde{a}_{k_m} \Bigr],
\;\;\; s=\{ q_r, m, q_z \}. \label{transf.phonon}
\end{multline}

The Hamiltonian $H$ is then given by the energy
(\ref{H_vor})-(\ref{H_int}) in terms of the variables
$\{\tilde{a}_k, \tilde{a}^{\dagger}_k\}, \{\tilde{c}_s,
\tilde{c}^{\dagger}_s\}$. Up to neglected relativistic
corrections, the variable transformation does not change the
spectrum of the elementary modes: the zero-order Hamiltonians are
given by (\ref{kelvons}), (\ref{phonons}) in terms of
$\{\tilde{a}_k, \tilde{a}^{\dagger}_k\}, \{\tilde{c}_s,
\tilde{c}^{\dagger}_s\}$. The transform (\ref{transf.phonon})
applied to (\ref{phonons}) generates the interaction term
\begin{multline}
H^{(\mathrm{rad})}_{\mathrm{int}} = \sum_{s,
\{k_i\}}(1-\delta_{m,0})\Bigl[\;\omega_s A_{s,k_1\ldots k_m} \;
\tilde{a}^{\dagger}_{k_1}\ldots
\tilde{a}^{\dagger}_{k_m}\tilde{c}^{\dagger}_s \\
+ \omega_s B_{s,k_1\ldots k_m} \; \tilde{a}_{k_1}\ldots
\tilde{a}_{k_m}\tilde{c}^{\dagger}_s\; \Bigr] + \mathrm{H.c}.
\label{H_radiation}
\end{multline}
Remarkably, the energy term $ \propto \int \! d^3r \, \eta
\bigl|\nabla \Phi_0\bigr|^2$ in (\ref{H_int}), which results in
the same operator structure as Eq.~(\ref{H_radiation}), is
irrelevant, being smaller in $\beta \ll 1$. It can be checked
straightforwardly, that the term $\propto \int \! d^3 r \; \eta
\nabla \varphi \cdot \nabla \Phi_0$ in Eq.~(\ref{H_int}) gives
Fetter's amplitudes of the elastic and inelastic scattering of
phonons \cite{Fetter_scattering} (see, however, the remark at the
end of the paper). In addition, this term leads to a
macroscopically small splitting of the phonon spectrum due to the
superimposed fluid circulation.

A kelvon is known to carry a quantum of (negative) angular
momentum projection \cite{Epstein_Baym}. The interaction
(\ref{H_radiation}) explicitly conserves the angular momentum: a
real process of the emission of a phonon with the angular momentum
$(-m)$ requires an annihilation of $m$ kelvons. Since
$\varepsilon_k \sim (a_0 k) \omega_k$, the total momentum
transferred to phonons in a radiation event should be small in
order to satisfy the energy conservation. Thus, the radiation by
one kelvon is kinematically suppressed. The leading radiation
process is the emission of the $m=-2$ (quadrupole) phonon mode,
the events involving more than two kelvons being suppressed by
$\alpha_k \ll 1$. First-order processes of the two-phonon emission
come from the term $\propto \int \! d^3 r \; \eta \nabla \varphi
\cdot \nabla \Phi_0$ of (\ref{H_int}). The amplitude of these
processes is suppressed by the relativistic parameter $\beta \ll
1$.

The Hamiltonian (\ref{H_radiation}) can be used to obtain the rate
of sound radiation by superfluid turbulence at zero temperature.
At large wave numbers, where the radiation is appreciable,
Kelvin-wave turbulence is characterized by $\alpha_k \ll 1$ and
features the spectrum $n_k = \langle a^{\dagger}_k a_k \rangle
\propto k^{-17/5}$ \cite{SvK_2004}. The occupation number decay
rate is $\dot{n}_k = - \sum_{s,k_1} W_{s,k,k_1}$, where
$W_{s,k,k_1}$ is the probability of the event $|0_s,\, n_k,\,
n_{k_1}\rangle \rightarrow |1_s,\, n_k\!-\!1, \, n_{k_1}\!-\!1 \,
\rangle$ per unit time. Applying the Fermi Golden Rule to
$W_{s,k,k_1}$ with the interaction (\ref{H_radiation}) and
replacing the sums by integrals, obtain
\begin{equation}
\dot{n}_k = - \frac{ (m_0 \kappa/2 \pi)^5 }{15 \pi  n  m_0} \,
\ln^5(1/a_0 k)\, (a_0 k)^5 \, k^5\, n_k^2. \label{kelvon_decay}
\end{equation}
Following Ref.~\cite{SvK_2004}, this formula allows us to obtain
the momentum scale $k_{\mathrm{ph}}$, at which the kelvon cascade
is cut off by the radiation of sound:
\begin{equation}
k_{\mathrm{ph}} \sim \frac{[a_0/R_0]^{6/31}}{\bigl[\ln
(R_0/a_0)\bigr]^{24/31}} a^{-1}_0, \label{k_ph}
\end{equation}
where $R_0$ is the typical distance between the vortex lines in
the tangle.

Finally, we comment on Vinen's estimate of the power radiated per
unit length of the vortex line, Eq.~(2.24) of
Ref.~\cite{Vinen2001}. The power at the momentum $k$ can be
determined by $\Pi_k= - \sum_{k' \sim k} \varepsilon_{k'}
\dot{n}_{k'}/L$, with $\dot{n}_{k'}$ given by
(\ref{kelvon_decay}). In Ref.~\cite{Vinen2001}, the retarded
potential method is employed giving the estimate $\Pi'_k \propto
b^2_k \propto n_k$, while $\Pi_k \propto n^2_k$. In the
quasi-particle language, $\Pi'_k$ implies that the radiation is
governed by the conversion of \textit{one} kelvon into a phonon.
Vinen argues that this process becomes allowed in superfluid
turbulence, where kelvons are actually superimposed on vortex
kinks of typical size $\sim R_0 \gg k^{-1}$, or, equivalently, the
kelvon coupling to a kink lifts the ban on single-kelvon radiative
processes by effectively removing the momentum conservation
constraint. We note that the probabilities of such elementary
events are likely to be suppressed \textit{exponentially}, as it
is generically the case, say, for soliton-phonon interactions
(see, e.g., \cite{soliton1} and references therein.) The processes
involving kelvon-kink coupling should contain an exponentially
small factor $\sim \exp(-R_0 k)$, which arises from the
convolution of the smooth kink profile with the oscillating kelvon
mode.

In the problem of phonon scattering from a vortex, the answer
depends on whether the vortex is pinned or free \cite{Sonin}. In
fact, the elastic scattering  matrix element of
Ref.~\cite{Fetter_scattering} corresponds to the pinned case only.
In the Hamiltonian formalism developed here, the difference is due
to an {\it additional} matrix element generated by the $|m|=1$
term of the interaction Hamiltonian (\ref{H_radiation}) in the
second order of perturbation theory. We are indebted to Edouard
Sonin for raising this question. Furthermore, a second-order
amplitude generated by a combination of the $|m|=2$ term with the
$|m|=1$ one is of the same order as the direct first-order
amplitude of inelastic phonon scattering. Thus, the result of
Ref.~\cite{Fetter_scattering} for inelastic scattering needs to be
corrected accordingly.

This work was supported by the National Science Foundation under
Grant No. PHY-0426881.

\end{document}